\documentclass[times,twocolumn,final]{MEDIMA_assets/elsarticle}

\usepackage{amsmath}
\usepackage{amssymb}
\usepackage{array} 
\usepackage[ruled,linesnumbered]{algorithm2e}
\usepackage{algpseudocode}
\usepackage{bm}
\usepackage{bbding}
\usepackage{booktabs}
\usepackage{colortbl}
\usepackage{xcolor}
\usepackage{caption}
\usepackage{cite}
\usepackage{color}
\usepackage{comment}
\usepackage{enumerate}
\usepackage{float}
\usepackage{graphicx}
\usepackage{mathrsfs}
\usepackage{subfigure}
\usepackage{textcomp}
\usepackage{url}
\usepackage{svg}

\usepackage{wrapfig}
\usepackage{MEDIMA_assets/medima}
\usepackage{framed,multirow}
\usepackage{latexsym}
\usepackage{hyperref}

\definecolor{newcolor}{rgb}{.8,.349,.1}

\journal{Medical Image Analysis}

\begin{document}

\verso{Shouyue Liu \textit{et~al.}}

\begin{frontmatter}

\title{Beyond the Eye: A Relational Model for Early Dementia Detection Using Retinal OCTA Images}


\author[1]{Shouyue {Liu}}
\author[2]{Ziyi {Zhang} }
\author[1]{Yuanyuan {Gu}}
\author[1]{Jinkui {Hao}}
\author[3]{Yonghuai {Liu}}
\author[4]{Huazhu {Fu}}
\author[1]{Xinyu {Guo}}
\author[5]{Hong {Song}\corref{cor1}} 
\author[2]{Shuting {Zhang}\corref{cor1}} 
\author[1]{Yitian {Zhao}\corref{cor1}} 

\cortext[cor1]{Corresponding authors: Yitian Zhao, Shuting Zhang, Hong Song; \\E-mail: yitian.zhao@nimte.ac.cn; shutingzhang@scu.edu.cn; songhong@bit.edu.cn} 

\address[1]{Ningbo Institute of Materials Technology and Engineering, Chinese Academy of Sciences, Ningbo, 315000, China}
\address[2]{Department of Neurology, West China Hospital, Sichuan University, Chengdu, 610000, China}
\address[3]{Department of Computer Science, Edge Hill University, Ormskirk, L39 4QP, United Kingdom}
\address[4]{Institute of High-Performance Computing, Agency for Science, Technology and Research, Singapore, 138632}
\address[5]{School of Computer Science and Technology, Beijing Institute of Technology, Beijing, 100081, China} 

\received{15 Apr 2024}
\finalform{13 Feb 2025}
\accepted{15 Feb 2025}
\availableonline{28 Feb 2025}

\begin{abstract}

Early detection of dementia, such as Alzheimer's disease (AD) or mild cognitive impairment (MCI), is essential to enable timely intervention and potential treatment. Accurate detection of AD/MCI is challenging due to the high complexity, cost, and often invasive nature of current diagnostic techniques, which limit their suitability for large-scale population screening. Given the shared embryological origins and physiological characteristics of the retina and brain, retinal imaging is emerging as a potentially rapid and cost-effective alternative for the identification of individuals with or at high risk of AD. In this paper, we present a novel PolarNet+ that uses retinal optical coherence tomography angiography (OCTA) to discriminate early-onset AD (EOAD) and MCI subjects from controls. 
Our method first maps OCTA images from Cartesian coordinates to polar coordinates, allowing approximate sub-region calculation to implement the clinician-friendly early treatment of diabetic retinopathy study (ETDRS) grid analysis. 
We then introduce a multi-view module to serialize and analyze the images along three dimensions for comprehensive, clinically useful information extraction. Finally, we abstract the sequence embedding into a graph, transforming the detection task into a general graph classification problem. A regional relationship module is applied after the multi-view module to explore the relationship between the sub-regions. Such regional relationship analyses validate known eye-brain links and reveal new discriminative patterns.
The proposed model is trained, tested, and validated on four retinal OCTA datasets, including 1,671 participants with AD, MCI, and healthy controls. Experimental results demonstrate the performance of our model in detecting AD and MCI with an AUC of 88.69\% and 88.02\%, respectively. Our results provide evidence that retinal OCTA imaging, coupled with artificial intelligence, may serve as a rapid and non-invasive approach for large-scale screening of AD and MCI. The code is available at \href{https://github.com/iMED-Lab/PolarNet-Plus-PyTorch}{https://github.com/iMED-Lab/PolarNet-Plus-PyTorch}, and the dataset is also available upon request. 

\end{abstract}
\begin{keyword}
 Alzheimer's Disease, OCTA images, Deep-learning, Polar Transformation.
\end{keyword}

\end{frontmatter}

\section{INTRODUCTION}

Alzheimer’s Disease (AD) is a progressive and irreversible neurodegenerative condition that is being diagnosed with increasing frequency and contributing substantially to the global disease burden~\citep{gaugler2021alzheimer}. Mild cognitive impairment (MCI) can be an early stage of memory or cognitive ability loss, and is an intermediate stage between cognitive normalcy and AD, with a high likelihood of regression to AD.
Early and accurate diagnosis of AD/MCI is critical to facilitate timely interventions and treatments. Currently, brain imaging, including magnetic resonance imaging (MRI) and positron emission tomography (PET), and neurobiological testing, such as cerebrospinal fluid amyloid, tau, and genetic risk scores, are commonly used in hospitals for the diagnosis of AD and MCI. However, they suffer from such limitations as being time-consuming, invasive, low accuracy, or high cost, which hinders their adoption in mass screening and routine clinical practice\citep{saykin2010alzheimer}.

The eye and brain share a similar tissue origin, and their similarity and association of structural characteristics and functional mechanisms have been previously investigated \citep{yin2024compartmentalized,liu2016association}.
Recently, significant differences in retinal biomarkers between AD and healthy controls have been reported, suggesting that AD affects the eyes and leads to changes in retinal structures. For example,
Wu et al.\citep{wu2020retinal} reported that both AD and MCI patients have a loss of retinal microvascular density in the macular region.
Curcio \citep{curcio2018viewing} extracted structural features in optical coherence tomography angiography (OCTA), and showed that cognitively impaired participants have a significant decrease in the thickness of inner fovea compared to healthy controls.
Zabel et al. \citep{zabel2019comparison} demonstrated that AD is associated with retinal neuronal apoptosis and retinal vascular dysfunction. 
These studies suggest that retinal microvascular attenuation may serve as a potential biomarker for signs of MCI and AD. Thus, retinal imaging has become a potential tool for detecting AD and MCI, and previous studies~\citep{9869953,cheung2022deep} have mainly applied machine learning methods over color fundus photography (CFP) for AD/MCI detection. 
Kim et al.~\citep{9869953} used CFP to train a modified MobileNet model to identify individuals with AD. This work modified the attention mechanism to the weighted attention mechanism and applied the mask-adding process to predict the likelihood of AD. 
Cheung et al. ~\citep{cheung2022deep} used thousands of CFP samples to develop a deep-learning model for the detection of AD. 
These methods rarely follow the clinical region-based analysis routine, which limits their ability to incorporate valuable clinical statistical findings and generate clinical-friendly results.

\begin{figure}[t]
    \centering
    \includegraphics[width=88mm]{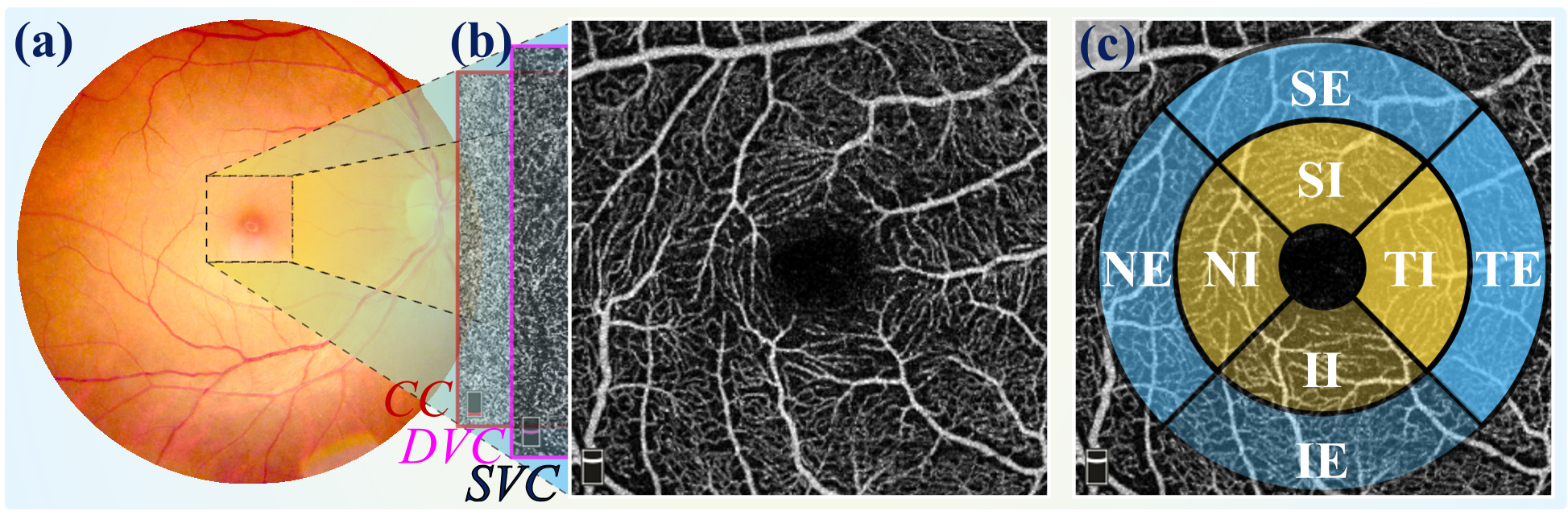}
    \caption{Demonstration of two different retinal imaging modalities of the same eye. (a) The CFP image and (b) its corresponding macula-centered OCTA images, such as superficial vascular complex (SVC), deep vascular complex (DVC), and choriocapillaris (CC). (c) The ETDRS grid applied on the OCTA image: temporal-inner (TI), temporal-external (TE), superior-inner (SI), superior-external (SE), nasal-inner (NI), nasal-external (NE), inferior-inner (II), and inferior-external (IE) sectors indicated.}
    \label{fundus_octa_polar_trans_fig}
\end{figure}

Most existing work has used the CFP imaging modality for the study of AD or MCI. Although CFP has advantages in terms of accessibility and cost-effectiveness, its native resolution (60-300 $\mu m$ in vessel diameter) is insufficient for imaging the retinal microvascular network in detail, as demonstrated in 
Fig.~\ref{fundus_octa_polar_trans_fig}-(a): this prevents the detection of subtle vascular changes in the early stages of diseases such as MCI. In contrast, Optical Coherence Tomography Angiography (OCTA) is an advanced imaging modality that provides non-invasive and rapid imaging of the retinal microvasculature, and choroidal capillaries with high resolution (5-6 $\mu m$ in diameter) across multiple layers, including the superficial vascular complex (SVC), deep vascular complex (DVC) and choriocapillaris (CC). Their maximum projection of OCTA flow signals, a.k.a. \textit{en face} images (Fig.~\ref{fundus_octa_polar_trans_fig}-(b)), which enhances depth-resolved microvascular imaging and facilitates the detection of subtle vascular changes, is critical for accurate screening and early diagnosis of many eye-related diseases, such as diabetic retinopathy \citep{sampson2022towards}.
In clinical practice, ophthalmologists often use regional analysis tools to study retinal biomarkers on retinal images. The most commonly used is the Early Treatment of Diabetic Retinopathy Study grid (ETDRS), as shown in 
Fig.~\ref{fundus_octa_polar_trans_fig}-(c). The ETDRS grid is a standardized grid that divides the retina into nine regions with three concentric circles and two orthogonal lines: a central foveal ring of 1 \textit{mm} diameter, an inner macular ring, and an outer macular ring. The ETDRS grid provides a systematic and consistent assessment of the macular region, allowing a more specific evaluation of retinal changes in a standardized manner, which can provide a more nuanced understanding of the disease\citep{rohlig2019enhanced,demirkaya2013effect,xu2018assessment}. As illustrated in 
Fig.~\ref{fundus_octa_polar_trans_fig}-(c), it can be found that both the ETDRS grid and the \textit{en face} images of the retina have circular characteristics that follow the nature of biology. Large vessels in SVC grow around a circle and gradually disappear near the center of the circle, forming a capillary-free foveal avascular zone (FAZ)\citep{conrath2005foveal}. Thus, in our previous work~\citep{xie2023deep}, we used the ETDRS grid to investigate the association of structural features with AD and MCI, and the results showed significant reductions in vessel area density and vessel length density in specific regions of the inner vascular complexes in AD and MCI participants.

Inspired by the above observations and findings, we propose a novel end-to-end framework in this work, namely PolarNet+, to take full advantage of clinical region-based analysis for EOAD and MCI detection using OCTA images. We aim to integrate the region-based feature extraction procedure, which is consistent with the ETDRS grid, into a deep learning-based classification model. To obtain a more accurate and clinical-acceptable result, it is worth noting that we specifically designed an approximate sector convolution, and the polar transformation was then applied to the regions of the ETDRS grid in order to take advantage of spatial constraints and improve the feature extraction and classification performances. 

\begin{figure*}[t]
\centering
\includegraphics[width=180mm]{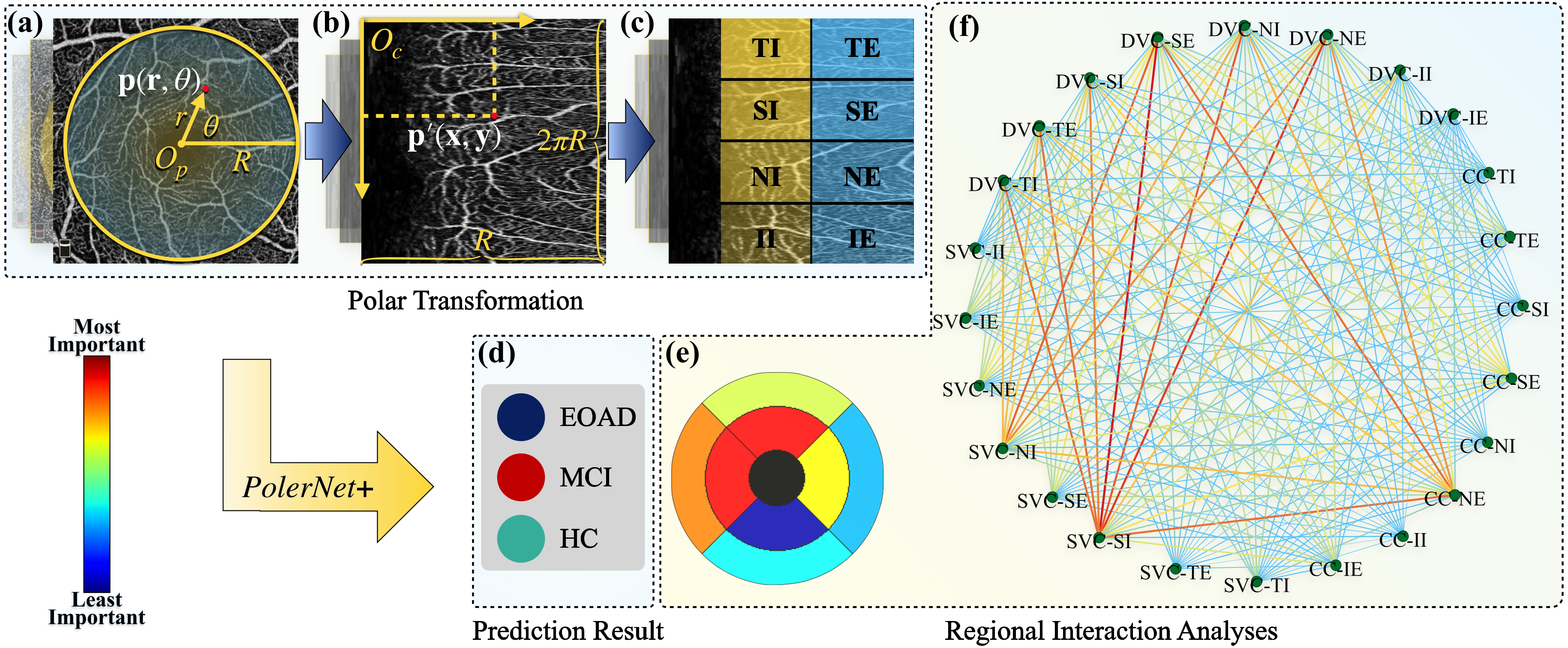}
\caption{A schematic illustration of the proposed PolarNet+ for EOAD/MCI detection using OCTA images and its regional-interaction analysis. (a)-(b) Illustrations of the polar and Cartesian coordinate systems, respectively. (c) The ETDRS grid in Fig. \ref{fundus_octa_polar_trans_fig}-(c) is applied to the OCTA image after polar transformation. (d) PolarNet+ categorizes the input OCTA images into EOAD, MCI, and healthy controls (HC). (e)-(f) Visualizations of importance maps and relationship graph. Different colors indicate different levels of significance. Annotations: ${O_p}$: the transformation center, also the FAZ center, also the pole of the polar coordinate system; ${O_c}$: the origin of the Cartesian coordinate system, where the horizontal axis is the ${X}$ axis, and the vertical axis is the ${Y}$ axis; ${R}$: the radius of the region of interest. }
\label{cfp_octa_pt_fl}
\end{figure*}

The proposed PolarNet+ significantly extends our work published in MICCAI-2023\citep{liu2023polar}, which was only verified on a single dataset containing 114 subjects. In this work, we expand the data pool from 114 to 1671 participants and make significant technical improvements. As a result, the new approach can also discriminate MCI subjects from healthy controls using retinal OCTA images, while achieving more accurate detection results. 
Overall, the major differences with the previous work and the main contributions of this work can be summarised as follows:

$\bullet$  We consider our work to be the first attempt in this research area for the detection of EOAD and MCI using retinal OCTA images. In addition to providing classification results, our model also provides a regional importance map and a regional relationship graph, highlighting discriminative patterns that drive decision-making, and revealing connections between retinal regions that are informative of neurological disease.

$\bullet$  We develop a 3-dimensional-serialization technique that models different retinal regions as sequences, allowing accurate extraction of image features. This approach captures spatial patterns that were previously overlooked by the CNN model, resulting in improved representational capabilities.

$\bullet$  We introduce a rewiring-based graph reasoning module to fully exploit and understand the relationships across diverse retinal and choroidal layers, leveraging the unique characteristics of OCTA data. This achieves promising classification performance. More importantly, it provides region connections to improve model transparency thereby facilitating a more clinically acceptable insight.

$\bullet$  Our clinical-friendly analyses validate known eye-brain links and reveal new discriminative patterns, demonstrating the model's potential as a computer-aided pathology tool for studying little-known associations between ophthalmic and complex neurological conditions.

\section{METHOD}
In this section, we detail the proposed PolarNet+ for EOAD/MCI detection method using retinal OCTA images, including image polar transformation, classification model architecture, and three specific modules for end-to-end training. 

Fig.~\ref{cfp_octa_pt_fl} illustrates the outline of our EOAD/MCI detection method using multiple \textit{en face} angiograms of OCTA as input. We first employ VAFF-Net\citep{hao2022retinal} to locate the FAZ center point on the SVC layer, and then transform the original images from polar coordinates to Cartesian coordinates, as shown in Fig.~\ref{cfp_octa_pt_fl} (a) and (b). 
These transformed images are fed into the PolarNet+ for the extraction of sequential features in circle-area, ring-area, and sector-area along three dimensions. After sequencing, the sequences that encode complementary region-specific information are treated as graph nodes and fed into a regional relationship module. Finally, PolarNet+ aggregates the node features and relationships for the generation of the final detection output. Furthermore, PolarNet+ is capable of generating two distinct visualisations, thereby facilitating a more clinically acceptable insight., as demonstrated in Fig.~\ref{cfp_octa_pt_fl} (e) and (f): a region importance map highlighting discriminative patterns that drive the decision-making, and a regional relationship graph revealing connections between retinal areas that are informative for neurological conditions.

\subsection{Polar transformation of OCTA image}

Here we introduce the polar transformation to realize region-based analysis. It aims to map coordinates from the polar coordinates $(r, \theta)$ to the Cartesian coordinate system $(x, y)$. 
Many image analysis tasks, such as pattern analysis, shape identification, or feature extraction, benefit from this transformation: e.g., Fu et al. \citep{fu2018joint} use polar transformation to convert the radial relationship to a spatial relationship in the image segmentation task.

As shown in Fig.~\ref{cfp_octa_pt_fl} (a) and (b), the polar transformation converts the region of interest (yellow circle) into a Cartesian coordinate system with the FAZ center ${O_p(x_o, y_o)}$ defined as the pole point. The original image is represented as points in the polar system $p(r, \theta)$: their corresponding points in the Cartesian system are represented by $p'(x, y)$ with the horizontal axis as the ${X}$ axis, and the vertical axis as the ${Y}$ axis. 
The following equations give the relationship between these two coordinate systems:
\begin{equation}\small
\left\{\begin{array} { l } 
{ x = r \operatorname { cos } \theta } \\
{ y = r \operatorname { sin } \theta }
\end{array} \Leftrightarrow \left\{\begin{array}{l}
r=\sqrt{x^{2}+y^{2}} \\
\theta=\tan ^{-1} y / x
\end{array}\right.\right. \   
\end{equation}
In order to preserve as much of the original data as possible, we chose the largest internally connected circle as the region of interest and retained the outermost pixels of the region of interest as the edges were cropped out. The width of the transformed image is equal to the radius ${R}$ of the yellow circle, and the height is ${2 \pi R}$. 
Since the corners are cropped, the outermost pixels of the region of interest are kept to preserve as much of the original information as possible, and the part near ${O_p}$ is filled by nearest neighbor interpolation. The polar transformation represents the original image in the polar coordinate system by pixel-wise mapping\citep{fu2018joint} and has the following properties:

\subsubsection{Approximate sector-shaped computing} The basic operations of deep learning rely mainly on linear algebra, such as convolution and linear operations, and include elements of rectangular or linear operations, such as vectors, matrices, and kernels\citep{aggarwal2020linear}. However, in the real world, many semantics are non-rectangular, e.g., retinal (circular or fan-shaped), which makes deep learning-based networks less optimal for OCTA-based image analyses. 
For the polar transformation, the mapping relationship is fixed, enabling us to approximate the sector convolution with a rectangular convolution kernel, thus facilitating its implementation. 

For the sake of simplicity and clarity, 
Fig.~\ref{cfp_octa_pt_fl}(c) 
illustrates the ETDRS grids mapped on the polar transformed image. Approximate sector calculation functions can be adapted to any sector herein. For instance, when we perform a rectangular kernel convolution along the ${TI \rightarrow SI}$ direction on the transformed image, it is equivalent to performing a sector-shaped kernel convolution around the center of the FAZ in the original image in a counterclockwise direction.

\subsubsection{Equivalent augmentation} Since the transformation is a pixel-by-pixel mapping, applying data augmentation to the original image is equivalent to applying data augmentation in the polar system \citep{fu2018joint}. 
For example, we can implement the drift cropping operation in a polar coordinate  system by changing  the transformation center ${O_p(x_o, y_o)}$ and the start angle. This is equivalent to changing the transform radius ${R}$ and applying different cropping factors for data augmentation.

\begin{figure*}[t]
\centering
\includegraphics[width=180mm]{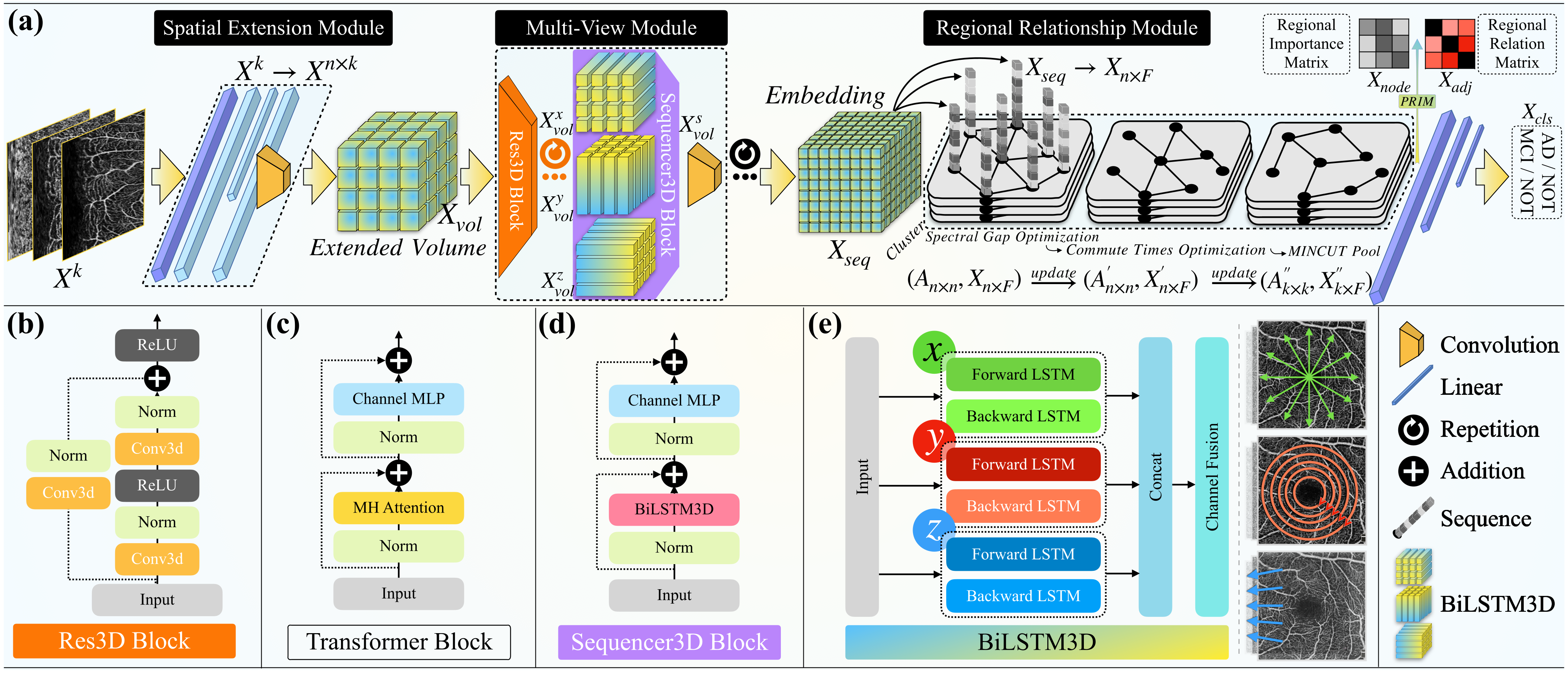}
\caption{The details of the proposed PolarNet+ (a) and its modules (b), (d) and (e). PolarNet+ comprises the spatial extension module, the multi-view module, and the regional relationship Module. PolarNet+ treats stacked images as volumes, and all operations are in 3D. We obtain the Sequencer3D block (d) by replacing the multi-head (MH) attention module in the transformer block (c) with BiLSTM3D (e). 
The right side of the subfigure (e) is the schematic representation of the 3D serialization processing. From top to bottom are explanations of the detailed splitting and mapping relations in the three directions (along the radius, around the pole, and along the depth). Here, we use one-way arrows for easy understanding. As for the implementation, we use BiLSTM for comprehensive feature extraction.}
\label{network_detail_fig}
\end{figure*}

\subsection{The classification model: PolarNet+}

OCTA provides in-depth information to visualize the retinal and choroidal microvascular network and the structure of the FAZ at different layers. To this end, we take advantage of three different \textit{en face} angiograms from different layers and explore intra-instance and inter-instance relationships across different retinal and choroidal layers, exploiting the unique characteristics of OCTA data. In this work, the SVC, DVC, and CC layers are used as input to the given PolarNet+, and we stack these three \textit{en face} images as 3D volume data.

The overall architecture of the proposed PolarNet+ is shown in Fig. \ref{network_detail_fig}.  It consists of three individual components: the spatial extension module, the multi-view module, and the regional relationship module. Since the stacked images are considered volumetric data, all computational operations are three-dimensional.



\subsubsection{Spatial Extension Module (SEM)}

The spatial extension module is designed to solve the thin volume problem. Considering that we only stack three images as a volume, if the number of stacked layers of the input is too small (e.g., we input three layers), the features in the stacked dimension will be very short (3 for raw data and 1 after convolution with a kernel size of 3), and numerous (${Height \times Width}$ e.g., ${224\times224=50176}$) for raw data. This leads to insufficiently rich feature semantics, and also greatly reduces the speed of computation. 
SEM can map three input layers to eight or more using linear layers, and the convolution can then extract features from multiple channels. By applying SEM, the data is transformed from a wide and thin shape to a thick and relatively narrow volume format.


Denote the input OCTA image as $X^k \in X$ of size $h \times w$, $k \in [1, K]$, and K as the total number of images from one patient.
The spatial extension module consists of a multi-layer perceptron (MLP) layer, a convolution layer, and a ReLU activation layer, which can be represented as: 
\begin{equation}\small
X_{vol} = \operatorname{SEM}({X^k}) = \operatorname{ReLU}(\operatorname{Conv}(\operatorname{MLP}({X^k})))
\end{equation}
where the $X_{vol}$ is the spatial extended volume.

\subsubsection{Multi-View Module (MVM)}
The multi-view module is designed to serialize and integrate volumetric features from different dimensions.
A more comprehensive representation can be obtained by extracting features separately along the \textit{x}, \textit{y}, and \textit{z} axes. Each MVM consists of one or more $\mathbf{Res3D}$ blocks followed by a $\mathbf{Sequencer3D}$ block. The $\mathbf{Res3D}$ block performs 3D convolutions to extract features, while avoiding vanishing gradients through residual connections: it can be repeated to adjust depth as needed.
The $\mathbf{Sequencer3D}$ block is inspired by the Sequencer module~\citep{tatsunami2022sequencer}, which was the first to successfully apply bidirectional long short-term memory (BiLSTM) for image classification.
We adopt a similar design, but modify it to take advantage of polar transformations.
Specifically, the $\mathbf{Sequencer3D}$ block consists of one $\mathbf{BiLSTM3D}$ unit. This contains three standard BiLSTMs one for each of the three dimensions. Each BiLSTM unit contains two LSTM units in two directions, \textit{i.e.}, forward and backward. 
Consider an input series denoted as ${\overrightarrow{x}}$ and let ${\overleftarrow{x}}$ represent the rearranged version of ${\overrightarrow{x}}$ in reverse order. The outputs obtained by processing ${\overrightarrow{x}}$ and ${\overleftarrow{x}}$ with their respective LSTMs are referred to as ${\overrightarrow{x_{for}}}$ and ${\overleftarrow{x_{back}}}$. ${\overrightarrow{x_{back}}}$ is then rearranged in the original order of the output ${\overleftarrow{x_{for}}}$, and the output ${b}$ of the BiLSTM, $\operatorname{B}(\cdot)$, is derived as follows:
\begin{equation}\small
\left\{\begin{matrix}
 \\
 \\
\end{matrix}\right. \begin{array}{l}
\overrightarrow{x_{for}}, \overleftarrow{x_{back}}=\operatorname{LSTM}_{\text{for}}(\vec{x}), \operatorname{LSTM}_{\text {back}}(\overleftarrow{x}), \\
b=\operatorname{B}(x)=\text { concatenate }\left(\overrightarrow{x_{for}}, \overrightarrow{x_{back}}\right).
\end{array}
\end{equation}

We implemented BiLSTM3D $\operatorname{B}_{\text{3D}}(\cdot)$ by applying BiLSTM on three dimensions. 
Since $X_{vol}$ is in volume format, we can split $X_{vol}$ into $X_{vol}^{x}$, $X_{vol}^{y}$, and $X_{vol}^{z}$, according to $x$, $y$, and $z$ dimensions. $\operatorname{Fusion}_{\text{channel}}(\cdot)$ denotes a channel fusion operation, and $X_{vol}^{s}$ denotes the volume sequence from three directions:
\begin{equation}
\left\{\begin{matrix}
 \\
 \\
 \\
\end{matrix}\right.
\begin{array}{l}
\left [ X_{vol}^{x},X_{vol}^{y},X_{vol}^{z} \right ] ^{T}= \operatorname{embedding}(X_{vol}) \\
X_{vol}^{s}=\operatorname{concatenate}(\operatorname{B}(X_{vol}^{x}),\operatorname{B}(X_{vol}^{y}),\operatorname{B}(X_{vol}^{z})), \\
X_{vol}^{'} = \operatorname{B}_{\text{3D}}(X_{vol}^{s}) =\operatorname{Fusion}_{\text{channel}}(X_{vol}^{s}).\\
\end{array}
\end{equation}

Here, we get an updated ${X_{vol}}$ as ${X_{vol}^{'}}$. We define a recursive operation $\operatorname{\Gamma}_{t}(\operatorname{f}_{\text{i}}(\cdot),\lambda),t \in \mathbb{Z}^{+}$ which denotes $\operatorname{f}_{\text{t}}(...(\operatorname{f}_{2}(\operatorname{f}_{1}(\lambda))))$, where $\operatorname{f}_{\text{i}}(\cdot)$ is the function to recursion, and $\lambda$ is an input. We repeat $\mathbf{Res3D}$ block $\operatorname{Res}(\cdot)$ several times and tandem the blocks to a $\mathbf{Sequencer3D}$ block $\operatorname{Seq}(\cdot)$ to get a multi-view block $\operatorname{P}(\cdot)$. Then, by recursing multi-view blocks, we obtain multi-view module $\operatorname{P}_{\text{m}}(\cdot)$:
\begin{equation}\small
\left\{\begin{matrix}
 \\
 \\
 \\
 \\
 \\
\end{matrix}\right.
\begin{array}{l}  
\operatorname{S}_{1}(\lambda) = \operatorname{B}_{\text{3D}}(\operatorname{Norm}(\lambda))+\lambda, \\
\operatorname{Seq}(\lambda)= \operatorname{MLP}(\operatorname{Norm}(\operatorname{S}_{1}(\lambda)))+\operatorname{S}_{1}(\lambda), \\
\operatorname{P}(\lambda) = \operatorname{ReLU}(\operatorname{Conv}(\operatorname{Seq}(\Gamma_{t1}(\operatorname{Res}(\cdot),\lambda)))), \\
\operatorname{P}_{\text{m}}(\lambda) = \Gamma_{t2}(\operatorname{P}(\cdot),\lambda),t1,t2 \in \mathbb{Z}^{+},\\
X_{seq} = \operatorname{P}_{\text{m}}(X_{vol}^{'}).\\
\end{array}
\end{equation}
where ${\lambda}$ represents any input, ${\operatorname{S}_{1}(\cdot)}$ is a function to package the intermediate stage, the $\operatorname{Norm}(\cdot)$ is the normalisation function
and ${X_{seq}}$ is the processed sequences data.

\subsubsection{Regional Relationship Module (RRM)}
Modeling the relationships and connections between different regions in the ETDRS grid is critical to establishing the association between pathological conditions 
and the retinal microvasculature for the accurate detection of EOAD and MCI. 
To this end, we propose the regional relationship module (RRM), which transforms the image classification problem into a graph representation, with region sequences modeled as graph nodes. Graph neural networks are naturally suited to learning from relational data and can embed high-dimensional connectivity patterns in a low-dimensional space \citep{wu2020comprehensive}.
Critically, rather than relying on a predefined adjacency matrix, the RRM incorporates a rewiring mechanism \citep{arnaiz2022diffwire} to learn the edges between regions. 
Through rewiring, it can learn an enriched representation of the relationships \citep{nguyen2023revisiting} between different regions within the ETDRS grid, which then allows the revelation of critical interdependencies associated with disease.
The RRM contains two differentiable layers, the commute times layer and the spectral gap optimization layer, for rewiring. The commute times layer $\operatorname{CT}(\cdot)$ \citep{arnaiz2022diffwire} identifies salient edges based on resistance, while the spectral gap optimization $\operatorname{GAP}(\cdot)$ \citep{arnaiz2022diffwire} layer optimizes the topology using spectral graph theory.

To find optimal clusters in the graph, we introduce MinCUT pooling \citep{bianchi2020spectral}. 
Given an adjacency matrix $A \in R^{n \times n}$ and a feature matrix $X \in R^{n \times F}$, we obtain a node cluster $(A_{n \times n}, X_{n \times F})$ where $n$ denotes the number of nodes and $F$ denotes the dimensions of the features. 
For a given input node cluster $(A_{n \times n}, R_{n \times F})$, a MinCUT pooling layer $\operatorname{MC}(\cdot)$ is applied to learn a new number $k$ of clusters: $(A_{n \times n}, X_{n \times F}) \to (A_{k \times k}, X_{k \times F}), k < n$.

We denote an operation that applies a linear operation $L$ to a function $f(\cdot)$ as $f_{L}(\cdot)$, which can be represented as a regional relationship module $RRM(\cdot)$:
\begin{equation}\small
\left\{\begin{matrix}
 \\
 \\
 \\
 \\
\end{matrix}\right.
\begin{array}{l}
\operatorname{S_{2}}(\lambda) = \operatorname{GAP}_{\text{L}}(\operatorname{Linear}(\lambda)),\\
\operatorname{S_{3}}(\lambda)= \operatorname{Conv}(\operatorname{CT}_{\text{L}}(\lambda)),\\
\operatorname{RRM}(\lambda) = \operatorname{Conv}(\operatorname{MC}_{\text{L}}(\operatorname{S_{3}}(\operatorname{S_{2}}(\lambda))),\\
\left [ X_{cls},X_{node},X_{adj} \right ] ^{T} = \operatorname{RRM}(X_{seq}).\\
\end{array}
\end{equation}
where ${\lambda}$ represents any input, ${\operatorname{S_{2}}}$ and ${\operatorname{S_{3}}}$ are the packaging of the GAP layer and the commute times layer. The ${X_{cls}}$, ${X_{node}}$, and ${X_{adj}}$ are the classification results, nodes matrix, and adjacency matrix, respectively.

\subsubsection{Polar Regional Importance Module (PRIM)}
In order to obtain the importance of different regions for EOAD and MCI prediction, we designed a polar regional importance module (PRIM) to facilitate prediction visualization and explanation.
The regional importance matrix $L_{\text {RI}}^{c}$ is calculated as follows:
\begin{equation}\small
L_{\text {RI}}^{c}=\operatorname{AvgPool} [ \operatorname{ReLU} (\sum_{k} \alpha_{k}^{c} A^{k} )  ] .
\end{equation}
where the $L_{\text {RI}}^{c}$ is the regional importance (RI) on the feature map ${k}$ for class ${c}$. 
We denote ${a_{k}^{c}}$ as the gradient of the score for class ${c}$, with respect to the feature map activations ${A^{k}}$ of the last layer \citep{liu2023polar}.

\section{Datasets}

Two different datasets were used: ROAD (Retinal OCTA for EOAD study) and ROMCI (Retinal OCTA for MCI study). All OCTA images in these two datasets were obtained from a multi-center case-control study for the detection of AD and MCI. 
The SVC, DVC, and CC angiograms were used for training, validation, and testing of the proposed automated method, as they together provide a comprehensive representation of the vasculature from superficial to deep.
Images from any given subject were allocated only to either the training or test sets, to avoid information leakage. All images analyzed were fovea-centered. 
The clinical protocol of this study was approved by the Ethics Committee of the Cixi Institute of Biomedical Engineering, Chinese Academy of Sciences, and adhered to the tenets of the Declaration of Helsinki. Informed written consent was obtained from the participants enrolled in our study.

\subsection{Retinal OCTA data for EOAD study (ROAD)}

The ROAD dataset consists of an internal subset (ROAD-I) used for model development and internal testing and an external subset (ROAD-II) used exclusively for external testing. ROAD-I includes 810 OCTA volumes from the Affiliated People's Hospital of Ningbo University, China, featuring 199 early-onset AD subjects and 611 controls. 
Inclusion criteria for EOAD subjects adhered to the National Institute on Aging and Alzheimer's Association (NIA-AA) guidelines, with diagnosis occurring before the age of 65. Exclusion criteria ensured the absence of other brain disorders, substance abuse, suicidal behaviors, and neurological diseases. 
The internal test and training sets used a fivefold cross-validation approach, with 20\% of the data allocated for internal testing and 80\% for training. ROAD-II, an independent dataset acquired from another center (West China Hospital, Sichuan University, Chengdu, China). It included 382 OCTA volumes (150 EOAD and 232 controls). By using it we aim to further validate the generalisability of our model.

\subsection{Retinal OCTA data for MCI study (ROMCI)}

The ROMCI dataset also consists of internal and external subsets, ROMCI-I and ROMCI-II. The former includes 545 OCTA volumes (104 MCI and 441 controls) from the Second Affiliated Hospital of Zhejiang University, Hangzhou, China. Participants with MCI were clinically assessed and diagnosed according to the diagnostic guidelines and recommendations of the Petersen Criteria. Clinical history, cognitive testing, and neuroimaging were reviewed for accuracy by an experienced neurologist specializing in memory disorders.
Exclusion criteria included significant sensory impairment and psychiatric disorders. ROMCI-II includes 180 OCTA volumes (35 MCI and 145 controls) obtained from the Affiliated People's Hospital of Ningbo University, Ningbo, China, and follows the same inclusion criteria as ROMCI-I. 

\begin{table*}[!t]\small
\centering
\caption{EOAD detection performances over the ROAD-I and II datasets using different methods. The best performance is highlighted in boldface. }\label{tabroad}
\resizebox{1\textwidth}{!}{ 
\setlength{\tabcolsep}{3mm}{
\begin{tabular}{l|ccc|ccc} 
\toprule
\multirow{2}*{\textbf{Methods}} & \multicolumn{3}{c|}{\textbf{ROAD-I}} & \multicolumn{3}{c}{\textbf{ROAD-II}} \\
~ &   ACC (mean${\pm}$std) & AUC (mean${\pm}$std) & Kappa (mean${\pm}$std) &  ACC (mean${\pm}$std) & AUC (mean${\pm}$std) & Kappa (mean${\pm}$std) \\
\hline
ResNet-34\citep{he2016deep} 	& 0.8006${\pm}$0.0146 & 0.8097${\pm}$0.0143 & 0.5891${\pm}$0.0297 & 0.7852${\pm}$0.0160 & 0.8022${\pm}$0.0272 & 0.5540${\pm}$0.0377\\
EfficientNet-B3\citep{tan2019efficientnet} 	& 0.7787${\pm}$0.0154 & 0.7945${\pm}$0.0088 & 0.5426${\pm}$0.0285 & 0.6915${\pm}$0.0160 & 0.7080${\pm}$0.0311 & 0.3630${\pm}$0.0399 \\
ConvNeXt-S\citep{liu2022convnet} 	& 0.7820${\pm}$0.0062 & 0.7782${\pm}$0.0116 & 0.5485${\pm}$0.0165 & 0.7558${\pm}$0.0209 & 0.7684${\pm}$0.0138 & 0.4938${\pm}$0.0472 \\
${\operatorname{HorNet-S_{GF}}}$\citep{rao2022hornet} 	& 0.7754${\pm}$0.0142 & 0.7980${\pm}$0.0193 & 0.5393${\pm}$0.0298 & 0.7019${\pm}$0.0227 & 0.7090${\pm}$0.0197 & 0.3935${\pm}$0.0517 \\
VAN-B6\citep{guo2022visual} 	& 0.7832${\pm}$0.0181 & 0.7949${\pm}$0.0187 & 0.5537${\pm}$0.0364 & 0.7345${\pm}$0.0182 & 0.7415${\pm}$0.0138 & 0.4516${\pm}$0.0421 \\
ViT-Base\citep{kolesnikov2021image} 	& 0.6298${\pm}$0.0264 & 0.6329${\pm}$0.0078 & 0.2502${\pm}$0.0412 & 0.6120${\pm}$0.0198 & 0.6220${\pm}$0.0133 & 0.2195${\pm}$0.0413 \\
SwinV2-T\citep{liu2022swin} 	& 0.6616${\pm}$0.0151 & 0.6764${\pm}$0.0126 & 0.3119${\pm}$0.0271 & 0.6780${\pm}$0.0164 & 0.6924${\pm}$0.0248 & 0.3429${\pm}$0.0360 \\
Early fusion\citep{hermessi2021multimodal} 	& 0.8039${\pm}$0.0174 & 0.8182${\pm}$0.0096 & 0.5951${\pm}$0.0366 & 0.6822${\pm}$0.0138 & 0.6952${\pm}$0.0221 & 0.3438${\pm}$0.0315 \\
Middle fusion \citep{zhou2019deep} 	& 0.8105${\pm}$0.0146 & 0.8127${\pm}$0.0189 & 0.6043${\pm}$0.0287 & 0.7337${\pm}$0.0150 & 0.7483${\pm}$0.0250 & 0.4509${\pm}$0.0437 \\
Late fusion\citep{heisler2020ensemble} 	& 0.8160${\pm}$0.0113 & 0.8175${\pm}$0.0136 & 0.6229${\pm}$0.0218 & 0.7820${\pm}$0.0172 & 0.7894${\pm}$0.0129 & 0.5457${\pm}$0.0437  \\
MCC\citep{zhou20213d} 	& 0.7722${\pm}$0.0156 & 0.8032${\pm}$0.0218 & 0.5259${\pm}$0.0339 & 0.7326${\pm}$0.0122 & 0.7512${\pm}$0.0226 & 0.4491${\pm}$0.0340 \\
MUCO-Net\citep{wang2022screening} 	& 0.8314${\pm}$0.0170 & 0.8286${\pm}$0.0096 & 0.6490${\pm}$0.0296 & 0.7825${\pm}$0.0065 & 0.7927${\pm}$0.0140 & 0.5469${\pm}$0.0187 \\
PolarNet\citep{liu2023polar} 	& 0.8489${\pm}$0.0187 & 0.8505${\pm}$0.0137 & 0.6882${\pm}$0.0355 & 0.8160${\pm}$0.0126 & 0.8162${\pm}$0.0156 & 0.6147${\pm}$0.0261 \\
\hline
\rowcolor{yellow!30} PolarNet+ w/o RRM 	& 0.8533${\pm}$0.0169 & 0.8545${\pm}$0.0140 & 0.6939${\pm}$0.0352 & 0.8359${\pm}$0.0207 & 0.8412${\pm}$0.0220 & 0.6536${\pm}$0.0486 \\
 \rowcolor{yellow!60} {\bfseries PolarNet+} 	& {\bfseries 0.8839${\pm}$0.0122} & {\bfseries 0.8869${\pm}$0.0059} & {\bfseries 0.7565${\pm}$0.0235} & {\bfseries 0.8569${\pm}$0.0131} & {\bfseries 0.8655${\pm}$0.0091} & {\bfseries 0.6994${\pm}$0.0314} \\

\bottomrule
\end{tabular}}}
\end{table*}

\section{Experimental Results}

\subsection{Implementation details}

To mitigate the impact of data imbalance, we employed the following strategies. Firstly, we applied data augmentation techniques, including rotating the minority class images (±20 degrees), to increase the diversity and representation of EOAD and MCI samples. In addition, a class-weighted loss function was used during model training to ensure balanced attention to all classes, regardless of their frequency in the dataset. The same data augmentation procedures, including rotation by +/- 20 degrees, were applied consistently across all comparison methods to ensure a fair comparison. 

We selected SVC, DVC, and CC as the input images, and the polar transformation was applied to them. 
Because the ETDRS regions of two eyes are symmetrical, before the transformation, all the images were flipped from the left eye (OS) to the right eye (OD). The transformed images were scaled down to a width of 224 pixels. 
We performed a hyperparameter search for all methods involved in the comparison (initial learning rate: from 1e-5 to 2e-4, stepped by 3e-5; dropout rate: from 4e-2 to 2e-1, stepped by 2e-2), and applied the corresponding hyperparameters that yielded optimal performance. The batch size is fixed to 16 and the same data augmentation procedures were applied consistently across all methods to ensure a fair comparison.

We implemented our method with PyTorch. A server running Ubuntu 20.04 with two Nvidia RTX 3090 GPUs was used to train the model. An initial learning rate of 2e-5 and a batch size of 16 were used, with AdamW \citep{loshchilov2017decoupled} as the optimizer. Fivefold cross-validation was used to exploit the data and maximise reliability. It is worth noting that the proposed PolarNet+ was implemented by repeating the Res3D block and the Sequencer3D blocks twice each to avoid over-fitting during training. 


In order to validate the proposed PolarNet+, the following state-of-the-art approaches were selected for comparison in each study, including 1) seven well-known universal classification methods: ResNet\citep{he2016deep},
EfficientNet\citep{tan2019efficientnet},
ConvNeXt\citep{liu2022convnet},
HorNet\citep{rao2022hornet},
VAN\citep{guo2022visual},
ViT\citep{kolesnikov2021image}, and 
SwinV2\citep{liu2022swin}; 2) four fusion-based methods: 
Early fusion\citep{hermessi2021multimodal},
Middle fusion \citep{zhou2019deep}, 
Late fusion\citep{heisler2020ensemble}, and 
MCC\citep{zhou20213d}; and 3) two OCTA-based dementia detection methods:
MUCO-Net\citep{wang2022screening}, and
PolarNet\citep{liu2023polar}. For a fair comparison, all these methods use multiple \textit{en face} angiograms as input. 

\subsection{Detection results}
We evaluated the PolarNet+ for the detection of EOAD and MCI on two cohorts, with both the internal and external datasets (ROAD-I/II and ROMCI-I/II). 

\subsubsection{Performance of EOAD detection}

We first evaluated PolarNet+ for EOAD detection over the internal dataset ROAD-I, and the quantitative results are shown in Table~\ref{tabroad}. It can be observed that our PolarNet+ outperforms all compared methods in three evaluation metrics: ACC (0.8839), AUC (0.8869), and Kappa (0.7565) on the ROAD-I dataset. It is worth noting that our approach exhibits superior performance compared to a specially tailored projection fusion approach, such as MUCO-Net. This observation underscores the key role that the exploitation of inter-projection relationships plays in determining the quality of classification results and, at the same time demonstrates that our 3D-based extraction method can achieve a more comprehensive fusion.
Furthermore, PolarNet+ shows improved performance compared to its ablated version (PolarNet+ without RRM), which lacks the regional relationship module. 
The primary rationale behind this is the superior ability of the regional relationship module to effectively model and extract relationships between ETDRS-guided regions, and to exploit the interplay of correlations and complementary information across different projections, thereby enhancing its performance in classification tasks.

\textbf{External validation} 
In order to further verify the generalisability of the PolarNet+ model, we report the results of the external validation of ROAD-II, as shown in the three right-hand columns of Table~\ref{tabroad}.
It is noteworthy that PolarNet+ shows superior performance in all the evaluated metrics. However, performance is relatively poorer than that on ROAD-I, as expected.
In particular, PolarNet+ shows a significant improvement, with its AUC value increasing from 0.8162 to 0.8655, surpassing its predecessor PolarNet (also the most competitive competitor), indicating a steady improvement in discrimination.
In addition, PolarNet+ outperforms PolarNet by improving its Kappa value from 0.6147 to 0.6994, an impressive improvement of 8.47\%.
This shows that the proposed method is effective in extracting discriminative features to ensure reliable EOAD detection, and demonstrates the stability of the model on datasets from different clinical centers.

We believe that the RRM plays an important role here (as demonstrated by the ablation experiments with PolarNet+ without RRM), as it models the learned neighborhood relationships as a part of knowledge so as to exploit further the correlation and complementarity between different \textit{en face} images and their sectors, making the network more consistent across different datasets \citep{lakhotia2022polarfly}. 

\begin{table*}[!t]\small
\centering
\caption{MCI detection performances over the ROMCI-I and II datasets using different methods. The best performance is highlighted in boldface.}\label{tabromci}
\resizebox{1\textwidth}{!}{ 
\setlength{\tabcolsep}{3mm}{
\begin{tabular}{l|ccc|ccc} 
\toprule
\multirow{2}*{\textbf{Methods}} & \multicolumn{3}{c|}{\textbf{ROMCI-I}} & \multicolumn{3}{c}{\textbf{ROMCI-II}} \\
~ &   ACC (mean${\pm}$std) & AUC (mean${\pm}$std) & Kappa (mean${\pm}$std) &  ACC (mean${\pm}$std) & AUC (mean${\pm}$std) & Kappa (mean${\pm}$std) \\
\hline

ResNet-34\citep{he2016deep} 	& 0.7542${\pm}$0.0164 & 0.7605${\pm}$0.0180 & 0.4599${\pm}$0.0433	& 0.7370${\pm}$0.0249 & 0.7504${\pm}$0.0321 & 0.4070${\pm}$0.0619	\\
EfficientNet-B3\citep{tan2019efficientnet} 	& 0.7617${\pm}$0.0229 & 0.7859${\pm}$0.0060 & 0.4620${\pm}$0.0539	& 0.6991${\pm}$0.0292 & 0.7449${\pm}$0.0463 & 0.3448${\pm}$0.0829	\\
ConvNeXt-S\citep{liu2022convnet} 	& 0.7226${\pm}$0.0186 & 0.7734${\pm}$0.0306 & 0.3887${\pm}$0.0418	& 0.6442${\pm}$0.0254 & 0.7054${\pm}$0.0457 & 0.2368${\pm}$0.0412	\\
${\operatorname{HorNet-S_{GF}}}$\citep{rao2022hornet} 	& 0.6909${\pm}$0.0370 & 0.7301${\pm}$0.0340 & 0.3116${\pm}$0.0655	& 0.6750${\pm}$0.0315 & 0.7166${\pm}$0.0307 & 0.3159${\pm}$0.0697	\\
VAN-B6\citep{guo2022visual} 	& 0.6986${\pm}$0.0253 & 0.7633${\pm}$0.0418 & 0.3436${\pm}$0.0520	& 0.6933${\pm}$0.0228 & 0.7344${\pm}$0.0323 & 0.3342${\pm}$0.0611	\\
ViT-Base\citep{kolesnikov2021image} 	& 0.7226${\pm}$0.0160 & 0.7729${\pm}$0.0249 & 0.3761${\pm}$0.0358	& 0.6351${\pm}$0.0137 & 0.6959${\pm}$0.0333 & 0.2457${\pm}$0.0509	\\
SwinV2-T\citep{liu2022swin} 	& 0.7618${\pm}$0.0188 & 0.8146${\pm}$0.0347 & 0.4620${\pm}$0.0177	& 0.6736${\pm}$0.0275 & 0.6926${\pm}$0.0288 & 0.2907${\pm}$0.0705	\\
Early fusion\citep{hermessi2021multimodal} 	& 0.7691${\pm}$0.0320 & 0.8079${\pm}$0.0193 & 0.4796${\pm}$0.0693	& 0.5539${\pm}$0.0286 & 0.6030${\pm}$0.0337 & 0.1133${\pm}$0.0644	\\
Middle fusion \citep{zhou2019deep} 	& 0.7784${\pm}$0.0182 & 0.7929${\pm}$0.0200 & 0.4864${\pm}$0.0561	& 0.5862${\pm}$0.0169 & 0.6390${\pm}$0.0362 & 0.1689${\pm}$0.0447	\\
Late fusion\citep{heisler2020ensemble} 	& 0.7760${\pm}$0.0179 & 0.8084${\pm}$0.0124 & 0.4807${\pm}$0.0452	& 0.5832${\pm}$0.0135 & 0.6458${\pm}$0.0501 & 0.1463${\pm}$0.0341	\\
MCC\citep{zhou20213d} 	& 0.7859${\pm}$0.0179 & 0.8383${\pm}$0.0174 & 0.5169${\pm}$0.0285	& 0.7281${\pm}$0.0262 & 0.7460${\pm}$0.0238 & 0.3958${\pm}$0.0623	\\
MUCO-Net\citep{wang2022screening} 	& 0.7978${\pm}$0.0229 & 0.8259${\pm}$0.0156 & 0.5384${\pm}$0.0480	& 0.7636${\pm}$0.0155 & 0.7964${\pm}$0.0312 & 0.4669${\pm}$0.0511	\\
PolarNet\citep{liu2023polar} 	& 0.8080${\pm}$0.0249 & 0.8372${\pm}$0.0388 & 0.5554${\pm}$0.0496	& 0.7770${\pm}$0.0362 & 0.7947${\pm}$0.0209 & 0.4917${\pm}$0.0691	\\
\hline
\rowcolor{yellow!30} PolarNet+ w/o RRM 	& 0.8296${\pm}$0.0251 & 0.8465${\pm}$0.0259 & 0.5945${\pm}$0.0525	& 0.7890${\pm}$0.0206 & 0.8156${\pm}$0.0248 & 0.4998${\pm}$0.0649	\\
\rowcolor{yellow!60} {\bfseries PolarNet+} 	& {\bfseries 0.8393${\pm}$0.0124} & {\bfseries 0.8802${\pm}$0.0141} & {\bfseries 0.6210${\pm}$0.0245}	& {\bfseries 0.8077${\pm}$0.0179} & {\bfseries 0.8316${\pm}$0.0224} & {\bfseries 0.5448${\pm}$0.0512}	\\

\bottomrule
\end{tabular}}}
\end{table*}

\subsubsection{Performance of MCI detection}

Table~\ref{tabromci} reports the MCI detection performances of different methods.
Overall, PolarNet+ gives the best results in terms of different evaluation metrics, with an ACC of 0.8393, an AUC of 0.8802, and a Kappa score of 0.6210. Notably, the OCTA images in the ROAD and ROMCI datasets were acquired using different OCTA systems. This difference allows an assessment of the generalization ability of the model across different machine configurations and imaging styles. In addition, it is important to note that the ROMCI dataset contains a smaller set of data compared to the ROAD dataset: this  makes the training of the model more susceptible to potential overfitting problems.

\textbf{External validation} 
As expected, all the methods had relatively low metric scores over the ROMCI-II compared to those over the ROMCI-I.
As in the EOAD detection task, the proposed PolarNet+ outperformed all competitors, achieving a superior ACC (0.8077), AUC (0.8316), and Kappa (0.5448). However, the performance gap between our MCI detection and other methods is narrower, when compared to the EOAD detection. This is probably due to the small amount of data, and the fact that there are fewer features for the detection of the disease in MCI. Overall, the results confirm the effectiveness and broad applicability of our model in detecting MCI in datasets obtained from different clinical centers.

\begin{figure}[H]
\centering
\includegraphics[width=88mm]{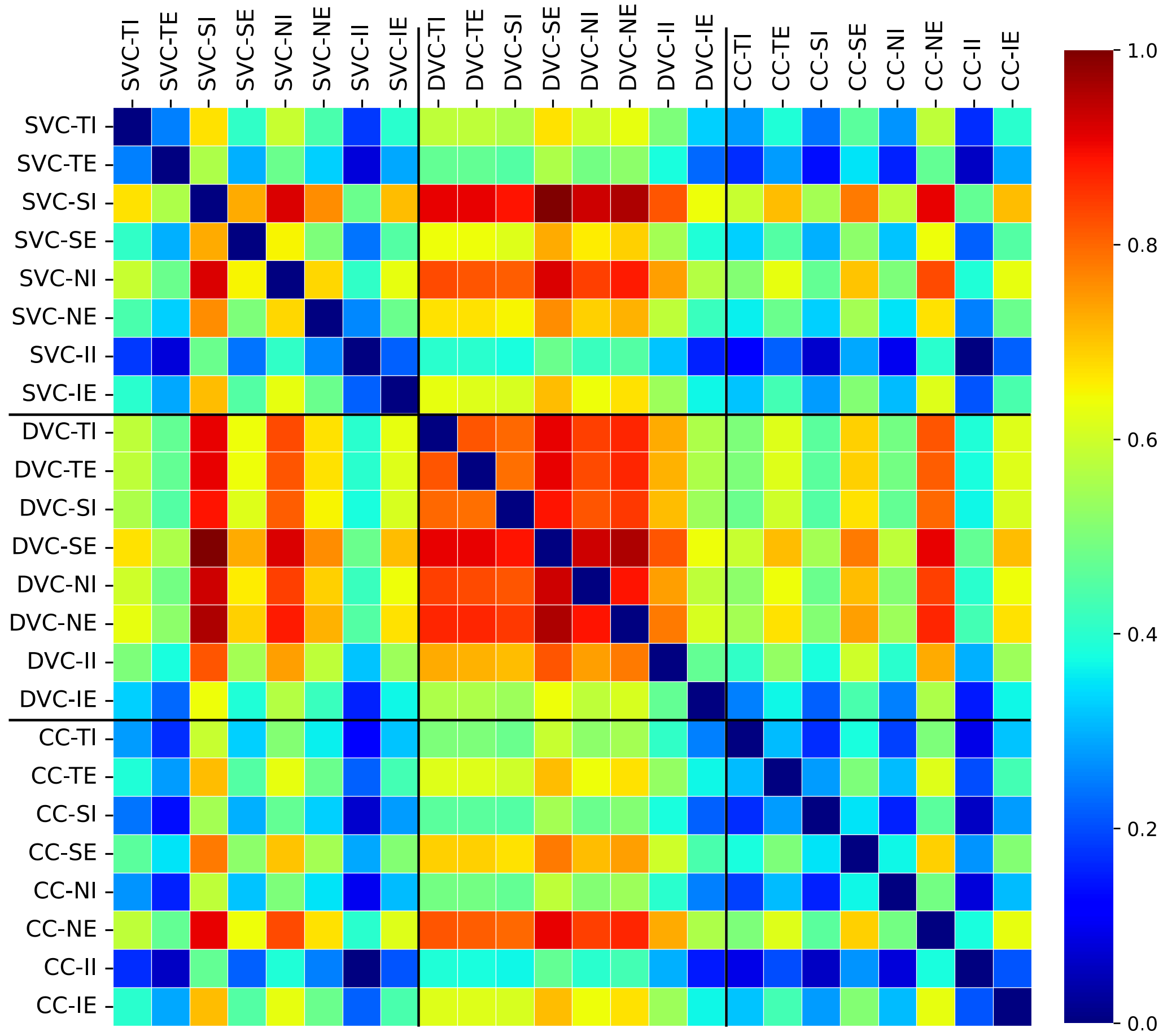}
\caption{Visualization of regional relationships by the means of regional adjacency matrix, generated from dataset ROAD-I.}
\label{region_relation_adj_matrix}
\end{figure}

\subsection{Explainability analysis} 

\begin{figure}
\centering
\includegraphics[width=83mm]{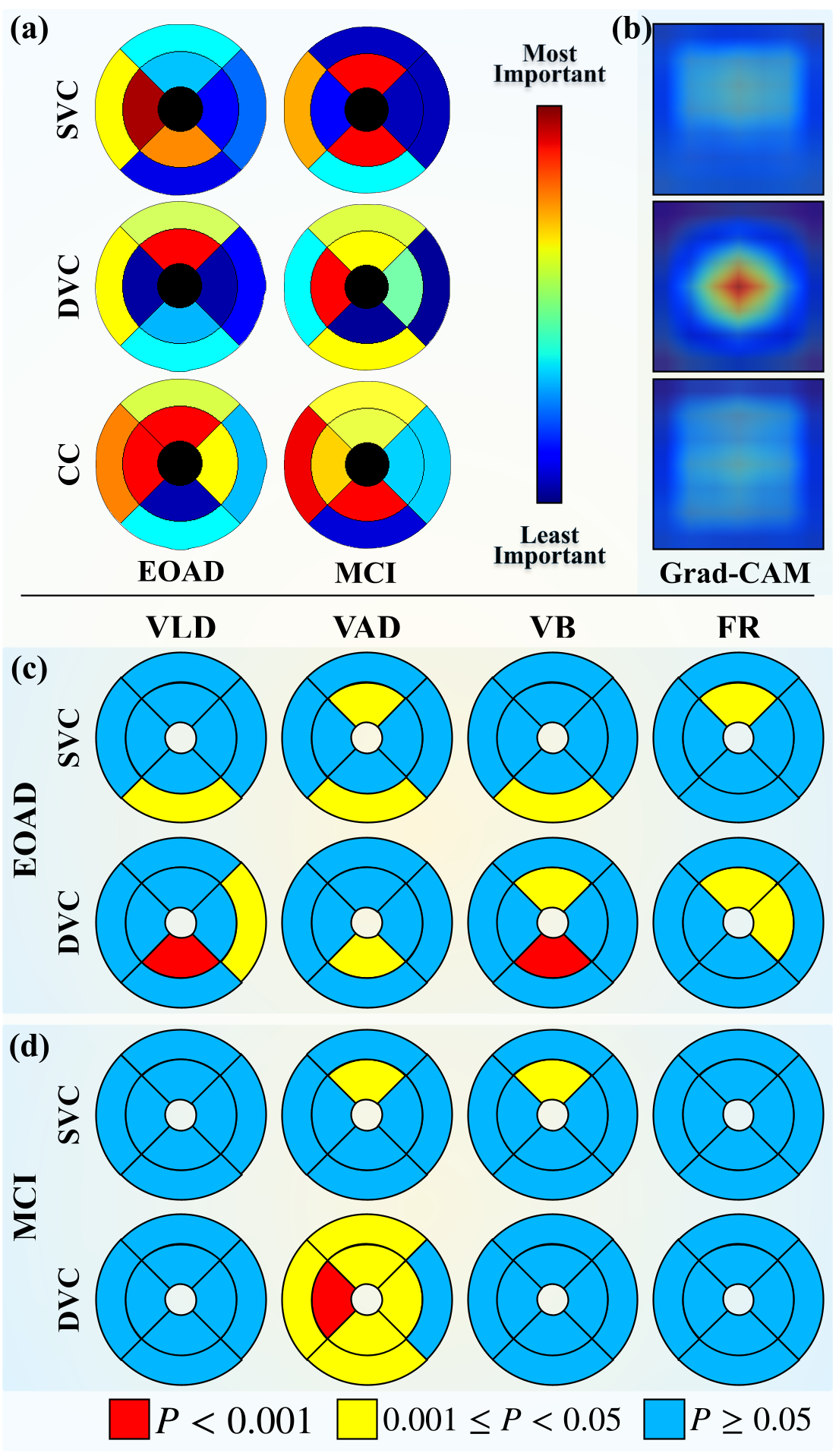}
\caption{Visualization of regional importance maps (a), generated from dataset ROAD-I and ROMCI-I. For comparison, we generated (b) the results of the normal visualization method, Grad-CAM, in the AD detection task on ROAD-I from a ResNet-18. (c) and (d) are the outcomes of the regional statistical assessments of the parameters generated from dataset ROAD-I and ROMCI-I, adjusted for covariates including age, gender, hypertension, diabetes, and education level, which were acquired through the utilization of the generalized estimation equation. }
\label{region_importance}
\vspace{-0.3cm}
\end{figure}

\begin{figure*}
\centering
\includegraphics[width=178mm]{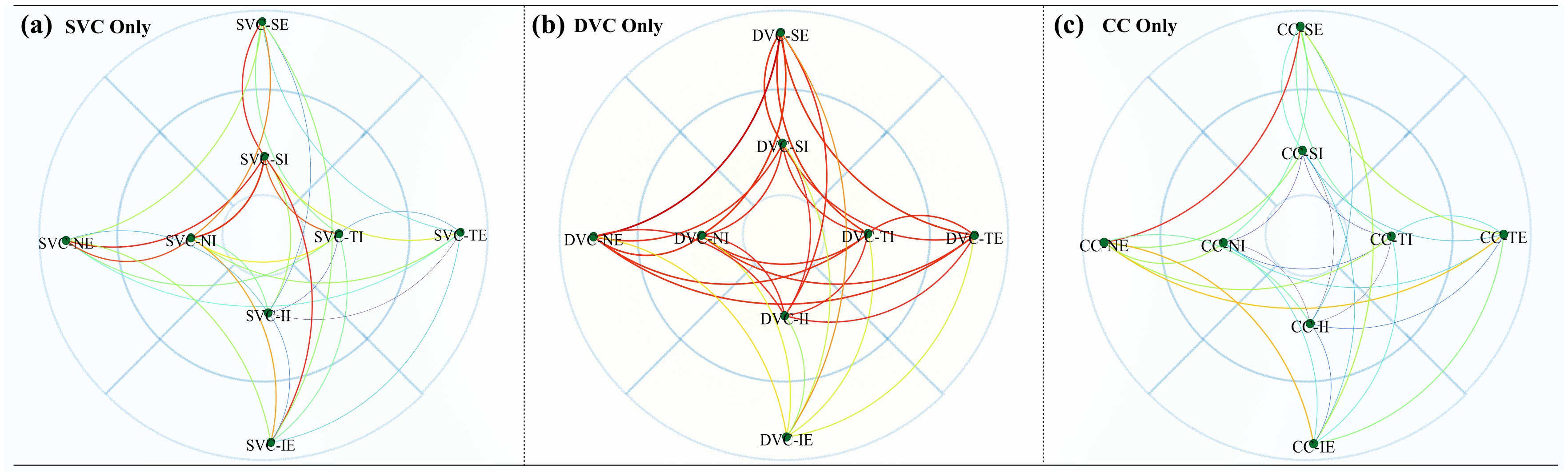}
\caption{Visualization of the local regional relationships, generated from dataset ROAD-I.
We arrange the nodes as their locations in the ETDRS grid for easier observation.
The nodes represent regions. Edge colors represent the relation strength, as in the legend. The definition of the ETDRS regions is shown in Fig.~\ref{fundus_octa_polar_trans_fig}-(c).}
\label{region_relation}
\end{figure*}

This subsection presents a visualization of the unique patterns learned by our model. The purpose is to gain a better understanding of the decision-making process of PolarNet+ and to explore how the relationships between different retinal and choroidal layers support the detection of EOAD and MCI. As stated in Section II, PolarNet+ can generate a regional importance map and a regional relationship graph. This allows for the identification of discriminative patterns that influence decision-making and the relationships between retinal areas that are relevant to neurological conditions.

\subsubsection{Regional importance analysis}
In the testing stage, we engaged the PRIM and produced a ${4\times2\times3}$ importance matrix, and accumulated the entire testing set. Subsequently, we performed an inverse operation of the polar transformation to generate the corresponding importance map based on the ETDRS grid, as shown in Fig.~\ref{region_importance}-(a). 
 Furthermore, we analyzed eight distinct parameters that collectively described the retinal microvasculature and the FAZ. These parameters encompass vascular length density (VLD), vascular area density (VAD), vascular bifurcation number (VB), vascular fractal dimension (VFD), FAZ area (FA), FAZ circularity (FC), FAZ roundness (FR), and FAZ solidity (FS). Subsequently, we examined the disparities between the EOAD/MCI and control groups, with the findings presented in Fig.~\ref{region_importance}-(c). The statistical analysis has shown that the parameters of VLD, VAD, and VB showed significance in SVC-IE and DVC-II, and VAD and FR showed significance in SVC-SI (all p-values are less than 0.001.
Compared to a commonly used visualization method, such as Grad-CAM \citep{selvaraju2017grad}, as shown in Fig.~\ref{region_importance}-(b), our region importance map is more clinician-friendly and easier to understand.

As shown in Fig.~\ref{region_importance}-(a), it can be seen that across the range of the entire ETDRS field, the importance of CC is the highest, which aligns with research \citep{corradetti2024choriocapillaris}, which may indicate a substantial decrease in choriocapillaris flow density among individuals diagnosed with EOAD. 
Simultaneously, the importance attributed to the DVC corresponds with research \citep{xie2023deep} discoveries and may indicate a substantial reduction in vascular area density and other pertinent factors within the DVC in EOAD (similar to Xie's \citep{xie2023deep} result of the vascular bifurcation number (VBN) analysis).
Interestingly, across the three layers, the higher importance occurred on the nasal sides and the superior sides, which may be explained by the research result from Asanad et al. \citep{asanad2019retinal}. Significant variations in the retinal choroid were observed both in terms of layers and regions, specifically in the nasal and temporal aspects concerning the optic nerve. 
Our results reveal a consistent pattern wherein heightened importance coincides with regions featuring greater micro-vessel density, such as the CC, DVC, and the parafovea, which is similar to the observation from studies \citep{corradetti2024choriocapillaris}. This may indicate a significant decrease in microvasculature in both the brain and retina of individuals affected by EOAD.

Additionally, our findings may also be confirmed partly through our statistical parameters results. As presented in Fig.~\ref{region_importance}-(c), we found that in EOAD, compared to healthy control, the parameters of VLD, VAD, and VB showed significance in SVC-IE and DVC-II, and VAD and FR showed significance in SVC-SI. For MCI, parameters of VAD and VB showed significance in SVC-SI, and parameters of VAD showed significance in nearly the entire DVC, except for DVC-TE. Many of the corresponding significant regions are presented in Fig.~\ref{region_importance}-(a), such as SVC-SI, DVC-SI, and DVC-NI.
Although our method works well, there are many high-importance regions that show differences with existing results that we could not explain. 
We should point out that the disparities observed may potentially stem from the network's capacity to unveil high-dimensional features that may not have been clinically identified to date, or to bias from the datasets.


\subsubsection{Regional relationship analysis}

The regional relationships graph is a new feature brought by PolarNet+, which can help researchers generate a new understanding of the relations of retinal vascular partitions. 
We visualize the adjacency matrix generated by the network, using Gephi \citep{bastian2009gephi}, in the form of a graph of the regional relationship, as shown in Fig.~\ref{region_relation}, based on the ETDRS grid. Meanwhile, we use a heat map to visualize the adjacency matrix, which aims to provide a different observation perspective, as shown in Fig.~\ref{region_relation_adj_matrix}. From the region-based relations learned by the network in relation to EOAD, we could observe a number of relatively closely related regions by colorizing the weights.

From the Fig.~\ref{region_relation_adj_matrix}, it is evident that partitions DVC-SE and SVC-(NI, SI)\footnote{SVC-(NI, SI) denotes SVC-NI and SVC-SI, the following is the same.} are closely linked, not only to their internal-layer sectors, but also to the other layer areas: this illustrates the relatively significant cooperating effects of the changes in the corresponding regions in the effects caused by EOAD on the captured retina. 
Meanwhile, some areas, such as DVC-IE and CC-(II, NI, SI), have relatively weak relations with the others, which indicates that the effect of these regions may be independent, even if their regional importance makes sense.


Upon examining the matrix heat map (Fig.~\ref{region_relation_adj_matrix}), we observed that the regions within the DVC exhibit stronger inter-regional connectivity compared to the regions in other retinal layers. Nearly all DVC regions demonstrate significant relationships with the regions in the SVC, particularly in the superior and nasal inferior quadrants, as well as with the regions in the CC, particularly in the NE quadrant. This is likely due to the anatomical positioning of the DVC between the SVC and CC within the retinal structure, naturally facilitating interactions with both adjacent layers. Furthermore, the DVC contains larger blood vessels and capillaries, which provide rich structural information regarding retinal vasculature, thereby offering more discriminative patterns for EOAD detection, as suggested by prior studies \citep{al2023innovative}.

A detailed examination of the matrix heat map (Fig.~\ref{region_relation_adj_matrix}) reveals that the regions within the DVC demonstrate a greater degree of inter-regional connectivity in comparison to the regions observed in other retinal layers. The majority of regions in the DVC layer exhibit notable correlations with those in the SVC, particularly in the superior and nasal inferior quadrants, as well as with regions in the CC, particularly in the NE quadrant. This may be attributed to the anatomical positioning of the DVC within the retinal structure, situated between the SVC and the CC. This natural configuration enables interaction with the adjacent retinal layers.  Moreover, the DVC contains a greater density of blood vessels and capillaries, which provide detailed structural information regarding the retinal vasculature. This offers a greater number of discriminative patterns for the detection of EOAD, as previously suggested by a clinical study \citep{al2023innovative}. 

A joint observation of Fig.~\ref{region_importance}-(a) and Fig.~\ref{region_relation} revealed that the DVC-(SI, NI), SVC-(SI, NI), and CC-(NE) exhibited higher importance weights and a stronger relationships than the other variables. 
These findings are consistent with those of a previous study~\citep{koronyo2023retinal}, which identified notable increases in amyloid ${\beta}$-protein (A${\beta}$) forms and novel intraneuronal A${\beta}$ oligomers on the nasal and superior sides.


\subsection{Effectiveness of polar transformation}
To evaluate the effectiveness of the polar transformation, we also performed the classification of different models with and without the polar transformation, and these results are presented in Table \ref{tabablationpolar}. 

It is evident that the majority of state-of-the-art models have demonstrated enhanced detection performance following the application of the polar transformation of the ETDRS grid, as evidenced by the improvement across all evaluation metrics. In comparison, our PolarNet+ demonstrates a more substantial improvement in performance. While the spatial positional features may undergo alterations following transformation, the key vascular features pertinent to our objective remain largely preserved within the ETDRS grid, and the improvement could be attributed to the transformation's ability to normalize vessel orientation and spatial relationships. 

Furthermore, the vascular alterations resulting from retinal disease are captured by the designed approximate sector convolution (polar transformation), which is intended to emulate the decision-making process frequently employed in clinical practice.  This shows that the model focuses on global and local patterns rather than specific spatial relationships, which explains its robustness to such distortions.

The results presented in Table \ref{tabablationpolar} indicate that the classification performance is not affected by the cropping of pixels. This finding aligns with the clinical studies \citep{hao2024early,zhang2019parafoveal,yeh2022retinal}, which identified that the pathological changes associated with AD/MCI are predominantly located within the parafoveal region with a specific radius.

\begin{table}[!t]
\centering
\caption{The detection performances of different models before and after polar transformation (Trans) over the ROAD-I dataset}\label{tabablationpolar}
\setlength{\tabcolsep}{3mm}{
\resizebox{0.48\textwidth}{!}{
\begin{tabular}{lcccc} 
\toprule
Methods & Trans & ACC (mean${\pm}$std) & AUC (mean${\pm}$std) & Kappa (mean${\pm}$std)\\
\toprule
\multirow{2}*{ResNet-34\citep{he2016deep}} & ${N}$ & 0.8006${\pm}$0.0146 & 0.8097${\pm}$0.0143 & 0.5891${\pm}$0.0297\\
~ & $\bm{Y}$ & {\bfseries 0.8313${\pm}$0.0088 } & {\bfseries 0.8293${\pm}$0.0188 } & {\bfseries 0.6491${\pm}$0.0238 } \\

\hline
\multirow{2}*{EfficientNet-B3\citep{tan2019efficientnet}} & ${N}$ & 0.7787${\pm}$0.0154 & 0.7945${\pm}$0.0088 & 0.5426${\pm}$0.0285\\
~ & $\bm{Y}$  & {\bfseries 0.8039${\pm}$0.0194 } & {\bfseries 0.7992${\pm}$0.0088 } & {\bfseries 0.5983${\pm}$0.0383 } \\

\hline
\multirow{2}*{ConvNeXt-S\citep{liu2022convnet}} & ${N}$ & 0.7820${\pm}$0.0062 & 0.7782${\pm}$0.0116 & 0.5485${\pm}$0.0165\\
~ & $\bm{Y}$  & {\bfseries 0.8050${\pm}$0.0285 } & {\bfseries 0.7884${\pm}$0.0169 } & {\bfseries 0.5994${\pm}$0.0605 } \\

\hline
\multirow{2}*{${\operatorname{HorNet-S_{GF}}}$\citep{rao2022hornet}} & ${N}$ & 0.7754${\pm}$0.0142 & 0.7980${\pm}$0.0193 & 0.5393${\pm}$0.0298 \\
~ & $\bm{Y}$  & {\bfseries 0.7985${\pm}$0.0126 } & {\bfseries 0.7991${\pm}$0.0078 } & {\bfseries 0.5844${\pm}$0.0285 } \\

\hline
\multirow{2}*{VAN-B6\citep{guo2022visual}} & ${N}$ & 0.7832${\pm}$0.0181 & 0.7949${\pm}$0.0187 & 0.5537${\pm}$0.0364\\
~ & $\bm{Y}$  & {\bfseries 0.8149${\pm}$0.0124 } & {\bfseries 0.8148${\pm}$0.0122 } & {\bfseries 0.6163${\pm}$0.0254 } \\

\hline
\multirow{2}*{ViT-Base\citep{kolesnikov2021image}} & ${N}$ & 0.6298${\pm}$0.0264 & 0.6329${\pm}$0.0078 & 0.2502${\pm}$0.0412\\
~ & $\bm{Y}$  & {\bfseries 0.6484${\pm}$0.0159 } & {\bfseries 0.6784${\pm}$0.0218 } & {\bfseries 0.2890${\pm}$0.0314 } \\

\hline
\multirow{2}*{SwinV2-T\citep{liu2022swin}} & ${N}$ & 0.6616${\pm}$0.0151 & 0.6764${\pm}$0.0126 & 0.3119${\pm}$0.0271\\
~ & $\bm{Y}$  & {\bfseries 0.6933${\pm}$0.0106 } & {\bfseries 0.7118${\pm}$0.0192 } & {\bfseries 0.3799${\pm}$0.0195 } \\

\hline
\multirow{2}*{Early fusion\citep{hermessi2021multimodal}} & ${N}$ & 0.8039${\pm}$0.0174 & 0.8182${\pm}$0.0096 & 0.5951${\pm}$0.0366\\
~ & $\bm{Y}$  & {\bfseries 0.8368${\pm}$0.0179 } & {\bfseries 0.8364${\pm}$0.0137 } & {\bfseries 0.6592${\pm}$0.0341 } \\

\hline
\multirow{2}*{Middle fusion \citep{zhou2019deep}} & ${N}$ & 0.8105${\pm}$0.0146 & 0.8127${\pm}$0.0189 & 0.6043${\pm}$0.0287\\
~ & $\bm{Y}$  & {\bfseries 0.8456${\pm}$0.0243 } & {\bfseries 0.8358${\pm}$0.0133 } & {\bfseries 0.6811${\pm}$0.0482 } \\

\hline
\multirow{2}*{Late fusion\citep{heisler2020ensemble}} & ${N}$ & 0.8160${\pm}$0.0113 & 0.8175${\pm}$0.0136 & 0.6229${\pm}$0.0218\\
~ & $\bm{Y}$  & {\bfseries 0.8477${\pm}$0.0179 } & {\bfseries 0.8390${\pm}$0.0130 } & {\bfseries 0.6831${\pm}$0.0353 } \\

\hline
\multirow{2}*{MCC\citep{zhou20213d}} & ${N}$ & 0.7722${\pm}$0.0156 & 0.8032${\pm}$0.0218 & 0.5259${\pm}$0.0339\\
~ & $\bm{Y}$  & {\bfseries 0.7952${\pm}$0.0179 } & {\bfseries 0.7852${\pm}$0.0083 } & {\bfseries 0.5790${\pm}$0.0348 } \\

\hline
\multirow{2}*{MUCO-Net\citep{wang2022screening}} & ${N}$ & 0.8314${\pm}$0.0170 & 0.8286${\pm}$0.0096 & 0.6490${\pm}$0.0296\\
~ & $\bm{Y}$  & {\bfseries 0.8412${\pm}$0.0093} & {\bfseries 0.8462${\pm}$0.0136} & {\bfseries 0.6676${\pm}$0.0174} \\

\hline
\multirow{2}*{PolarNet\citep{liu2023polar}} & ${N}$ & 0.8171${\pm}$0.0113 & 0.8289${\pm}$0.0095 & 0.6253${\pm}$0.0211 \\
~ & $\bm{Y}$ & {\bfseries 0.8489$\pm$0.0187} & {\bfseries 0.8505$\pm$0.0137} & {\bfseries 0.6882$\pm$0.0355}\\

\hline
\multirow{2}*{PolarNet+} & ${N}$ & 0.8532${\pm}$0.0088 & 0.8605${\pm}$0.0055 & 0.6959${\pm}$0.0165 \\
~ & $\bm{Y}$ & {\bfseries 0.8839${\pm}$0.0122} & {\bfseries 0.8869${\pm}$0.0059} & {\bfseries 0.7565${\pm}$0.0235}\\

\bottomrule
\end{tabular}}}
\end{table}

\subsection{Model parameters analysis}
As shown in Table~\ref{tabparams}, we conducted a series of experiments to evaluate the performance of the named models by feeding them the same data (${4\times3\times224\times224}$). 

In our implementation for the EOAD/MCI detection problem, PolarNet+ is constructed by repeating the Res3D block and the multi-view module twice respectively. Consequently,  the depth of our implementation is relatively shallow, so its size is tiny (3.92M parameters) when compared to the popular universal models. 
Although the multi-view property and the relationship-analyzing property, which enhance the high detection performance, increase the computational load, modern computing architectures, fortunately, have already covered such scaled computational demands.
The advantage of the number of parameters highlights the promise of PolarNet+ deployment on low-cost devices, such as those used for community screening, which could increase the possibility of recognizing EOAD/MCI early.

\begin{table}[!t]
\centering
\caption{The parameter size of the different methods.}
\label{tabparams}

\setlength{\tabcolsep}{3mm}{
\resizebox{0.41\textwidth}{!}{
\begin{tabular}{l|r} 
\toprule
Methods	&	Parameter size (M)\\
\hline
ResNet-34\citep{he2016deep} 	&	21.80	\\
EfficientNet-B3\citep{tan2019efficientnet} 	&	10.70	\\
ConvNeXt-S\citep{liu2022convnet} 	&	49.46	\\
${\operatorname{HorNet-S_{GF}}}$\citep{rao2022hornet} 	&	49.63	\\
VAN-B6\citep{guo2022visual} 	&	199.06	\\
ViT-Base\citep{kolesnikov2021image} 	&	85.80	\\
SwinV2-T\citep{liu2022swin} 	&	27.58	\\
Early fusion\citep{hermessi2021multimodal} 	&	21.98	\\
Middle fusion \citep{zhou2019deep} 	&	64.55	\\
Late fusion\citep{heisler2020ensemble} 	&	65.93	\\
MCC\citep{zhou20213d} 	&	94.37	\\
MUCO-Net\citep{wang2022screening} 	&	60.18	\\
PolarNet\citep{liu2023polar} 	&	26.08	\\
\rowcolor{yellow!60}  {\bfseries PolarNet+} 	&	{\bfseries 3.92}	\\
\bottomrule
\end{tabular}}
}
\end{table}

\section{DISCUSSIONS}

To further examine the potential clinical-acceptability of PolarNet+ and to obtain more detailed information about the regional relationships of the OCTA ${en face}$ layers, we performed more detailed analyses, such as single-layer analysis (shown in Fig.~\ref{region_relation}-(a)-(c).) and inter-layer analysis.

When we focus on the SVC-DVC section in Fig.~\ref{region_relation_adj_matrix}, we can see that SVC-(SI, NI) connect to DVC-Tx\footnote{DVC-Tx denotes DVC-TI and DVC-TE, the following is the same.} and DVC-(Sx, Nx), where SVC-SI performs most actively. 
Turning to the DVC-CC section in Fig.~\ref{region_relation_adj_matrix}, all DVC regions except DVC-Ix have relations with CC-NE, among which the DVC-(SE, Nx) occupy a prominent position. Although SVC and CC are not directly connected to each other at the tissue structure level, some potential connections do exist: the aforementioned active regions SVC-(SI, NI) show a strong connection to CC-(SE, NE), as shown in the SVC-CC section in Fig.~\ref{region_relation_adj_matrix}. 
To summarize the inter-layer patterns, we can conclude that the SVC-(SI, NI), CC-NE, and the DVC (except DVC-Ix), play a key role in cooperating effects, and these regions are located on the nasal sides and the superior sides, which is almost in line with the pattern observed in our regional importance analysis. 

When we observe the single layer, inner-layer regional relationship patterns are clear. 
Fig.~\ref{region_relation}-(a) shows that in the SVC, there is only one strong connection between SVC-SI and SVC-NI, which is even weaker than those in the CC, Fig.~\ref{region_relation}-(c), where CC-NE is connected as a bridge to the CC-(SE, TE, IE), concentrating on the nasal side. This may suggest that the EOAD influences CC more strongly than the SVC. 
In the inner-DVC, Fig.~\ref{region_relation}-(b), connections between regions become the strongest and densest, concentrating on the superior sides, with the DVC-SE in the leading position, connecting to DVC-(Nx, Tx, SI). 

Our findings suggest a more robust connection between modifications in the DVC and EOAD compared to shifts in either the SVC or the CC. 
The DVC, characterized by capillaries with a slender cross-section, demonstrates heightened sensitivity to the advancement of the disease cascade \citep{wang2018emerging}. Therefore, we believe that the DVC is more important for the detection of EOAD.

\subsection{Limitation of this work}

We also acknowledge the limitations of this study.
We have conducted additional experiments in which the AD, MCI, and normal datasets were merged and evaluated as a multi-class classification problem. The results were as follows: ACC=0.8858 and AUC=0.9350. However, it was observed that the performance of a three-class classification was superior to that of a binary classification. A detailed examination revealed that the underlying cause was data heterogeneity. The OCTA data collected in two cohorts (EOAD and MCI) were obtained using two different OCTA devices. The model can readily distinguish between EOAD and MCI subjects based on the image style, rather than the pathological features observed in OCTA images. This finding motivates the development of a transfer learning model to normalise images collected from disparate cohorts or clinical centres in future research.

It is also difficult at this stage to validate the link between the explored regional relationships and disease pathogenesis. However, the main focus of this work is to develop an end-to-end detection model and to explore regional interactions in retinal OCTA images in the context of neurodegenerative diseases, which is arguably the first attempt in this field of research. The aim is to move from the analysis of individual features to a more comprehensive relational perspective. It is believed that this approach can help or guide clinicians and biologists to focus on relevant anatomical regions and understand the mechanisms of their association, thus providing a new avenue to deepen the understanding of disease mechanisms.

\section{CONCLUSION}

 In clinical practice, the region-based analysis technique is frequently employed to study OCTA image biomarkers and to understand the correlation with various eye-related diseases. This is typically achieved by utilising a region-based analysis technique such as the ETDRS grid. To this end, our framework integrates the ETDRS and gradCAM, and maps the model's decision-making to a region-based representation that aligns with clinical practice. This paper takes one step in addressing the critical EOAD/MCI detection issue through innovative retinal imaging and deep learning methods. PolarNet+ inherits its predecessor's ability to compute regional significance and provides views of regional relationships, with which we can obtain a more open and comprehensive view of AD/MCI analysis based on OCTA images.

 Our method extends beyond the mere averaging of data, incorporating regional relationship modelling to elucidate the manner in which different regions interact and contribute collectively to the decision-making process. This offers a more comprehensive view of the model's predictions, improving the  clinically acceptability using the assessment of regional importance. To the best of our knowledge, this work is the first attempt to generate a regional map of significant importance, specifically designed for EOAD/MCI  using OCTA images. This task-specific adaptation not only adds clinical relevance but also provides a new level of insight into regional disease patterns. Although gradCAM is a well-established method for visual explanations, averaging outputs within regions is necessary for tasks that rely on region-level analysis. This approach guarantees consistency with the clinical partitioning strategy and enables the model to concentrate on region-specific disease-relevant patterns.

Dementia remains a formidable global health challenge, necessitating early and accurate diagnosis for effective interventions. Our research builds upon the fundamental connection between the brain and the eye, capitalizing on the region-based method and physiological traits to explore retinal biomarkers for EOAD/MCI detection and classification. Retinal OCTA imaging overcomes many defects of established techniques in AD screening, as a non-invasive alternative, enabling the detection of subtle retinal vasculature changes that can serve as valuable AD biomarkers. In conclusion, the contributions of PolarNet+ are substantial, based on a new mapping paradigm for regional relationships, expanding the understanding of AD.
The journey from PolarNet to PolarNet+ represents a substantial advancement in the pursuit of clinical-friendly and precise analysis based on OCTA images. 

\section{Acknowledgment}This work is supported by the National Key Research and Development Program of China (2024YFB3815100), the National Science Foundation Program of China (62422122,  62272444), the Key Research and Development Program of Zhejiang Province (2020C03036, 2023C04017, 2024C03101),  Ningbo 2025 S\&T Mega project (2021Z134, 2022Z127, 2022Z134), the Key Project of Ningbo Public Welfare Science and Technology (2023S012).

\bibliographystyle{MEDIMA_assets/model2-names.bst}\biboptions{authoryear}
\bibliography{biblib.bib}

\end{document}